\documentclass[utf8]{FrontiersinHarvard} 

\usepackage{url,hyperref,lineno,microtype,subcaption,multirow,makecell,graphicx}
\usepackage[onehalfspacing]{setspace}

\def\keyFont{\fontsize{8}{11}\helveticabold }
\def\firstAuthorLast{Duricic {et~al.}} 
\def\Authors{Tomislav Duricic\,$^{1,2,*}$, Dominik Kowald\,$^{1,2,*}$, Emanuel Lacic\,$^{3}$, and Elisabeth Lex\,$^{1,*}$}
\def\Address{$^{1}$Institute of Interactive Systems and Data Science, Graz University of Technology, Graz, Austria \\
$^{2}$Know-Center, Graz, Austria  \\
$^{3}$Infobip, Zagreb, Croatia}

\def\corrAuthor{Tomislav Duricic, Dominik Kowald, Elisabeth Lex}
\def\corrEmail{tduricic@\{know-center.at, tugraz.at\}, dkowald@know-center.at, elisabeth.lex@tugraz.at}

\usepackage{array}
\usepackage{booktabs}
\usepackage{multirow}
\usepackage{caption}

\begin{document}
\onecolumn
\firstpage{1}

\title[Diversity, Serendipity and Fairness in GNN-based Recommender Systems]{Beyond-Accuracy: A Review on Diversity, Serendipity and Fairness in Recommender Systems Based on Graph Neural Networks} 

\author[\firstAuthorLast ]{\Authors} 
\address{} 
\correspondance{} 

\extraAuth{}

\maketitle

\begin{abstract}


By providing personalized suggestions to users, recommender systems have become essential to numerous online platforms. Collaborative filtering, particularly graph-based approaches using Graph Neural Networks (GNNs), have demonstrated great results in terms of recommendation accuracy. However, accuracy may not always be the most important criterion for evaluating recommender systems' performance, since beyond-accuracy aspects such as recommendation diversity, serendipity, and fairness can strongly influence user engagement and satisfaction. This review paper focuses on addressing these dimensions in GNN-based recommender systems, going beyond the conventional accuracy-centric perspective. We begin by reviewing recent developments in approaches that improve not only the accuracy-diversity trade-off but also promote serendipity and fairness in GNN-based recommender systems. We discuss different stages of model development including data preprocessing, graph construction, embedding initialization, propagation layers, embedding fusion, score computation, and training methodologies. Furthermore, we present a look into the practical difficulties encountered in assuring diversity, serendipity, and fairness, while retaining high accuracy. Finally, we discuss potential future research directions for developing more robust GNN-based recommender systems that go beyond the unidimensional perspective of focusing solely on accuracy. This review aims to provide researchers and practitioners with an in-depth understanding of the multifaceted issues that arise when designing GNN-based recommender systems, setting our work apart by offering a comprehensive exploration of beyond-accuracy dimensions. 

\tiny
 \keyFont{ \section{Keywords:} Review, Survey, Recommender Systems, Graph Neural Networks, Beyond Accuracy, Diversity, Serendipity, Fairness} 
\end{abstract}

\section{Introduction}

With their ability to provide personalized suggestions, recommender systems have become an integral part of numerous online platforms by helping users find relevant products and content~\cite{aggarwal2016recommender}. There are various methods employed to implement recommender systems, among which collaborative filtering (CF) has proven to be particularly effective due to its ability to leverage user-item interaction data to generate personalized recommendations~\cite{koren2021advances}. Recent advances in Graph Neural Networks (GNNs) have also had a significant impact on the field of recommender systems, and especially on collaborative filtering. GNN-based CF approaches have demonstrated exceptional results in terms of recommendation accuracy, which has traditionally been the main criterion for evaluating the performance of recommender systems~\cite{he2020lightgcn,pu2012evaluating}.

However, most studies have focused only on accuracy and have often neglected other equally or sometimes even more important aspects of recommender systems, such as diversity, serendipity, and fairness. The importance of these \textit{beyond-accuracy} dimensions is increasingly being recognized, as studies have shown that these aspects can have a significant impact on user satisfaction~\cite{abdollahpouri2019beyond}. For example, diverse and serendipitous recommendations can prevent the over-specialization of content and enhance user discovery. Fairness, on the other hand, ensures that the system does not discriminate against certain users or item providers, thereby promoting equitable user experiences~\cite{gao2023survey}. 

This review paper further explores these dimensions in the context of GNN-based recommender systems, going beyond the traditional accuracy-centric viewpoint. We discuss recent advances in approaches that not only improve the accuracy-diversity trade-off, but also promote serendipity and fairness. Furthermore, we highlight the practical issues encountered in assuring these dimensions when constructing GNN-based CF approaches, while preserving high recommendation accuracy. This review is intended to provide researchers and practitioners with a comprehensive understanding of the multifaceted optimization issues that arise when designing GNN-based recommender systems, thereby contributing to the development of more robust and user-centric recommender systems.

\section{Background}

Graph neural networks (GNNs) have recently emerged as an effective way to learn from graph-structured data by capturing complex patterns and relationships~\cite{hamilton2020graph}. Through the propagation and transformation of feature information among interconnected nodes in a graph, GNNs can effectively capture the local and global structure of the given graphs. Consequently, they emerge as an ideal method especially suitable for dealing with tasks involving interconnected, relational data such as social network analysis, molecular chemistry, and recommender systems among others.

In recommender systems, integrating Graph Neural Networks (GNNs) with traditional collaborative filtering techniques has been shown beneficial. Representing users and items as nodes in a graph with interactions acting as edges allows GNNs to provide more accurate personalized recommendations by discovering and utilizing intricate connections that would otherwise remain undetected~\cite{wang2019neural}. In particular, higher-order connectivity together with transitive relationships play an essential role when trying to extract user preferences in certain scenarios.

GNN-based recommender systems represent an evolving field with continuous advancements and innovations. Recent research has focused on multiple aspects of GNNs in recommender systems, ranging from optimizing propagation layers to effectively managing large-scale graphs and integration of auxiliary information~\cite{zhou2022graph}. Aside from these aspects, an expanding interest lies in exploring beyond-accuracy objectives for recommender systems. Such objectives include diversity, explainability/interpretability, fairness, serendipity/novelty, privacy/security, and robustness which offer a more comprehensive evaluation of the system's performance~\cite{wu2022graph,gao2023survey}. However, our work focuses primarily on three key aspects: diversity, serendipity, and fairness, since these aspects have a significant impact on user satisfaction, while also considering ethical concerns in the field of recommender systems. Ensuring diversity amongst recommendations minimizes over-specialization effects, benefiting users in product/content discovery and exploration~\cite{kunaver2017diversity}. Considering serendipity also helps to overcome the over-specialization problem by allowing the system to recommend novel, relevant, and unexpected items, thus improving user satisfaction~\cite{kaminskas2016diversity}. The aspect of fairness ensures that the system does not discriminate against certain users or item providers, thereby promoting equitable user experiences~\cite{deldjoo2023fairness}. 


Diversity, serendipity, and fairness in recommender systems are interconnected and often influence each other. For instance, increasing diversity can lead to more serendipitous recommendations, since users are exposed to a wider range of unexpected and less-known items~\cite{kotkov2020does}. Furthermore, focusing on diversity and serendipity can also promote fairness, since it ensures a more equitable distribution of recommendations across items and prevents the system from consistently suggesting only popular items~\cite{mansoury2020fairmatch}. However, it's important to note that these aspects need to be balanced with the system's accuracy and relevance to maintain user satisfaction. Considering beyond-accuracy dimensions contributes to supporting the development of GNN-based recommender systems that are not only robust and accurate but also user-centric and ethically considerate. 

While GNNs have seen rapid advancements, their application in recommender systems has also been the subject of several surveys. \cite{wu2022graph} and \cite{gao2023survey} provide a broad overview of GNN methods in recommender systems, touching upon aspects of diversity and fairness. \cite{dai2022comprehensive} delves into fairness in graph neural networks in general, briefly discussing fairness in GNN-based recommender systems. Meanwhile, \cite{fu2023deep} explores serendipity in deep learning recommender systems, with limited focus on GNN-based recommenders. Building on these insights, our review distinctively emphasizes the importance of diversity, serendipity, and fairness in GNN-based recommender systems, offering a deeper dive into these dimensions.

To conduct our review, we searched for literature on Google Scholar using keywords such as ``diversity", ``serendipity'', ``novelty'', ``fairness'', ``beyond-accuracy'', ``graph neural networks'' or ``recommender system''. We manually checked the resulting papers for their relevance and retrieved 21 publications overall from relevant journals and conferences in the field (see Table~\ref{table:summary}). While re-ranking and post-processing methods are often used when optimizing beyond-accuracy metrics in recommender systems~\cite{gao2023survey}, this paper specifically concentrates on advancements within GNN-based models, thus leaving these methods outside the discussion. Finally, it is important to highlight that diversity, serendipity, and fairness are extensively researched in recommender systems beyond GNNs. Broader literature across various architectures has provided insights into these challenges and their overarching solutions. While our paper primarily focuses on GNNs, we direct readers to consult these works for a comprehensive perspective~\cite{kaminskas2016diversity, wang2023survey}.

\section{Model Development}

The construction of a GNN-based recommender system is a complex, multi-stage process that requires careful planning and execution at each step. These stages include data preprocessing (DP), graph construction (GC), embedding initialization (EI), propagation layers (PL), embedding fusion (EF), score computation (SC), and training methodologies (TM). In this section, we provide an overview of this multi-stage process as it is crucial for understanding the specific stages at which current research has concentrated efforts to address the beyond-accuracy aspects of diversity, serendipity, and fairness in GNN-based recommender systems.

\begin{figure}[t!]
\centering
\includegraphics[width=\textwidth]{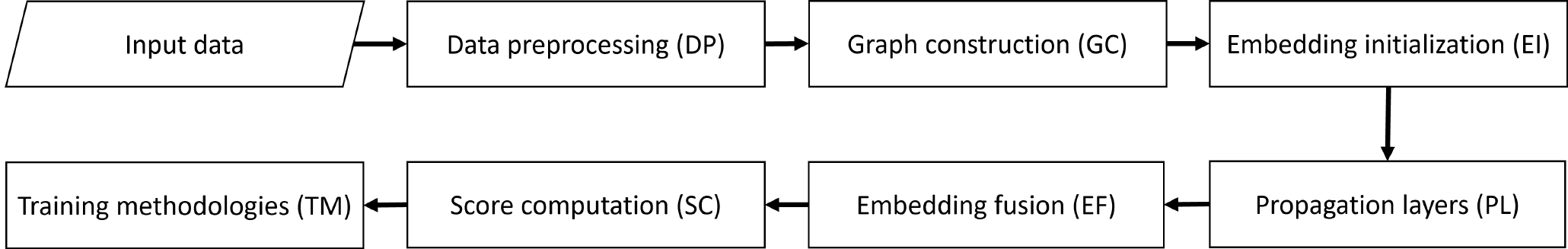}
\caption{The simplified multi-stage process of developing a GNN-based recommender system, each of these stages strongly impacts resulting recommendations and can be considered when designing a model that takes into account beyond-accuracy objectives.}
\label{fig:gnn-cf}
\end{figure}

\subsection{Data preprocessing, graph construction, embedding initialization}

The initial stage of developing a GNN-based collaborative filtering model is data preprocessing, where user-item interaction data and auxiliary information such as user/item features or social connections are collected and processed~\cite{lacic2015utilizing,duricic2018trust,fan2019graph,wang2019knowledge,duricic2020empirical}. Techniques like data imputation ensure that missing data is filled, providing a more complete dataset, while outlier detection helps in maintaining the data's integrity. Feature normalization ensures consistent data scales, enhancing model performance. Addressing the cold-start problem at this stage ensures that new users or items without sufficient interaction history can still receive meaningful recommendations~\cite{lacic2015tackling,liu2020heterogeneous}.

The graph construction stage is crucial, as the graph's structure directly influences the model's efficacy. Choosing the type of graph determines the nature of relationships between nodes. Adjusting edge weights can prioritize certain interactions while adding virtual nodes/edges can introduce auxiliary information to improve recommendation quality~\cite{wang2020disentangled,kim2022debiasing,wang2023collaboration}.

In the embedding initialization stage, nodes are assigned low-dimensional vectors or embeddings. The choice of embedding size balances computational efficiency and representation power. Different initialization methods offer trade-offs between convergence speed and stability. Including diverse information in the embeddings can capture richer user-item relationships, enhancing recommendation quality~\cite{wang2021pre}. This initialization can be represented as $H^{(0)} = \left[h_{\text{user}}^{(0)}; h_{\text{item}}^{(0)}\right]$, where $h_{\text{user}}^{(0)}$ and $h_{\text{item}}^{(0)}$ are the initial embeddings of the user and item nodes, respectively.

\subsection{Propagation layers, embedding fusion, score computation, training methodologies}

Propagation layers in GNNs aggregate and transform features of neighboring nodes to generate node embeddings, represented as $H^{(l+1)} = \sigma\left(D^{-1}A H^{(l)} W^{(l)}\right)$, where $H^{(l)}$ is the matrix of node features at layer $l$, $A$ is the adjacency matrix, $D$ is the degree matrix, $W^{(l)}$ is the weight matrix at layer $l$, and $\sigma$ is the activation function~\cite{hamilton2020graph}. There are numerous approaches built on this concept. For instance, \cite{he2020lightgcn} adopt a simplified approach, emphasizing straightforward neighborhood aggregation to enhance the quality of node embeddings; whereas \cite{fan2019graph} integrate user-item interactions with user-user and item-item relations, capturing complex interactions through a comprehensive graph structure.

Afterward, these embeddings are combined during the embedding fusion stage, forming a latent user-item representation used for score computation by applying a weighted summation, concatenation, or a more complex method of combining user and item embeddings~\cite{wang2019neural,he2020lightgcn}.

The score computation stage involves a scoring function to output a score for each user-item pair based on the fused embeddings. The scoring function can be as simple as a dot product between user and item embeddings, or it can be a more complex function that takes into account additional factors~\cite{wang2019neural,he2020lightgcn}.

Finally, in the training methodologies stage, a suitable loss function is selected, and an optimization algorithm, typically a variant of stochastic gradient descent, is used to update model parameters~\cite{rendle2012bpr,fan2019deep}.

Understanding the unique strengths of each stage outlined in this section is essential, and a comparative evaluation can guide the selection of the most suitable approach for specific collaborative filtering scenarios, such as addressing the challenges associated with beyond-accuracy metrics. In Table~\ref{table:summary}, we provide a comprehensive overview of existing literature, aiding readers in navigating the diverse methodologies and findings discussed throughout this review. 

\begin{table}[h!]
\renewcommand{\arraystretch}{1.3}
\centering
\caption{This table summarizes key literature on GNN-based recommender systems emphasizing beyond-accuracy metrics: Diversity, Serendipity, and Fairness. Each entry specifies the paper's publication venue/journal, targeted metric, a broad strategy categorization, and the model development stages the method utilizes or adapts to enhance the respective metric. These stages include data preprocessing (DP), graph construction (GC), embedding initialization (EI), propagation layers (PL), embedding fusion (EF), score computation (SC), and training methodologies (TM).}
\scalebox{0.76}{
\vspace{2mm}
\begin{tabular}{|c|c|c|c|}
\Xhline{3\arrayrulewidth}
\textbf{Beyond-accuracy} & \multirow{2}{*}{\textbf{Paper and venue/journal}} & \multirow{2}{*}{\textbf{Topic/contribution}} & \textbf{Model development stages} \\ 
\textbf{metric} &  &  & \textbf{utilized to tackle metric} \\\Xhline{3\arrayrulewidth} 
\multirow{17}{*}{Diversity} & \cite{isufi2021accuracy} & \multirow{3}{*}{Neighbor-based mechanism} & \multirow{3}{*}{GC, PL, EF, TM} \\ 
& Information Processing & & \\ 
& and Management & & \\ \cline{2-4}
& \cite{wang2020disentangled} & \multirow{2}{*}{Disentangling mechanisms} & \multirow{2}{*}{GC, EI, PL, EF, TM} \\ 
& ACM SIGKDD conf. & & \\ \cline{2-4}
& \cite{ye2021dynamic} & \multirow{2}{*}{Dynamic graph construction} & \multirow{2}{*}{EI, GC, TM} \\ 
& ACM RecSys conf.& & \\ \cline{2-4}
& \cite{yang2023dgrec} & \multirow{2}{*}{Neighbor-based mechanisms} & \multirow{2}{*}{PL, EF, TM} \\ 
& ACM WSDM conf. & & \\ \cline{2-4}
& \cite{zuo2023dtgcf} & \multirow{2}{*}{Adversarial learning} & \multirow{2}{*}{GC, PL, TM} \\ 
& MDPI Applied Sciences & & \\ \cline{2-4}
& \cite{ma2022contrastive} & \multirow{2}{*}{Contrastive learning} & \multirow{2}{*}{EI, GC, PL, TM} \\ 
& IEEE IJCNN conf. & & \\ \cline{2-4}
& \cite{zheng2021dgcn} & Neighbor-based mechanism,  & \multirow{2}{*}{PL, TM} \\ 
& ACM Web Conf.  & Adversarial learning  & \\ \cline{2-4}
& \cite{xie2021improving} & \multirow{2}{*}{Heterogeneous GNNs} &  \multirow{2}{*}{GC, PL, SC, TM}\\ 
& IEEE Trans. on Big Data & & \\ \Xhline{3\arrayrulewidth}
\multirow{11}{*}{Serendipity} & \cite{dhawan2022improvedgcn} & General GNN architecture & \multirow{2}{*}{-} \\ 
& Electronic Commerce & enhancements & \\ 
& Research and Applications & & \\ \cline{2-4}
& \cite{liu2020long} & \multirow{2}{*}{Long-tail recommendations} & \multirow{2}{*}{DP, EF, SC, TM} \\ 
& ACM RecSys conf. & &  \\ \cline{2-4}
& \cite{sun2020framework} & General GNN architecture &\multirow{2}{*}{GC, PL, EF, SC, TM} \\ 
& ACM SIGKDD conf. & enhancements & \\ \cline{2-4}
& \cite{zhao2022investigating} & \multirow{2}{*}{Normalization techniques} & \multirow{2}{*}{PL} \\ 
& ACM SIGIR conf. & & \\ \cline{2-4}
& \cite{boo2023serendipity}  & \multirow{2}{*}{Neighbor-based mechanisms} & \multirow{2}{*}{EI, EF, SC, TM} \\ 
& ACM IUI conf. & & \\ \Xhline{3\arrayrulewidth}
\multirow{17}{*}{Fairness} & \cite{xu2023fairness} & \multirow{2}{*}{Contrastive learning} & \multirow{2}{*}{GC, TM} \\ 
& Information Sciences & & \\ \cline{2-4}
& \cite{li2019long} & \multirow{2}{*}{Multimodal feature learning} & \multirow{2}{*}{GC, PL, EF} \\
& ACM CIKM conf. & & \\ \cline{2-4}
& \cite{liu2022self} & \multirow{2}{*}{Self-training mechanisms} & \multirow{2}{*}{PL, TM} \\ 
& Applied Soft Computing & & \\ \cline{2-4}
& \cite{kim2022debiasing} & \multirow{2}{*}{Neighbor-based mechanisms} & \multirow{2}{*}{PL, SC, TM} \\ 
& ACM CIKM conf. & & \\ \cline{2-4}
& \cite{yang2023debiased} & \multirow{2}{*}{Contrastive learning} & \multirow{2}{*}{GC, PL, EF, TM} \\ 
& ACM Web Conf. & & \\ \cline{2-4}
& \cite{wu2022equipping}  & \multirow{2}{*}{Neighbor-based mechanisms} & \multirow{2}{*}{GC, PL, EF, TM} \\ 
& ACM ASONAM conf. & & \\ \cline{2-4}
& \cite{gupta2019niser}  & \multirow{2}{*}{Long-tail recommendations} & \multirow{2}{*}{PL, SC, TM} \\ 
& ACM CIKM conf. & & \\ \cline{2-4}
& \cite{liu2022mitigating} & Neighbor-based mechanisms, & \multirow{2}{*}{GC, PL, TM} \\ 
& Neural Computing & Adversarial learning & \\ 
& and Applications & & \\ \Xhline{3\arrayrulewidth}
\end{tabular}
}
\label{table:summary}
\end{table}

\section{Diversity in GNN-based Recommender Systems}
\subsection{Definition and importance of diversity}

Diversity in recommender systems indicates how different the suggested items are to a user. It's vital for recommendation quality, preventing over-specialization, and boosting user discovery. Diverse recommendations offer users a wider item range, enhancing satisfaction and user engagement~\cite{kunaver2017diversity,duricic2021my}. Diversity has two types: intra-list (variety within one recommendation list) and inter-list (variety across lists for different users)~\cite{kaminskas2016diversity}.



\subsection{Review of recent developments in improving accuracy-diversity trade-off}

A number of innovative approaches have emerged recently to tackle recommendation diversity using graph neural networks (GNNs). These methods can be broadly categorized based on the specific mechanisms or strategies they employ:

\begin{itemize}
    \item \textbf{Neighbor-based mechanisms~\footnote{Neighbor-based mechanisms aggregate and propagate information from neighboring nodes (users or items) to enhance the representation of a target node, capturing intricate relational patterns for improved recommendations~\cite{wu2022graph}.}:} An approach introduced by~\cite{isufi2021accuracy} combines nearest neighbors (NN) and furthest neighbors (FN) with a joint convolutional framework. The \textit{DGRec} method diversifies embedding generation through submodular neighbor selection, layer attention, and loss reweighting~\cite{yang2023dgrec}. Additionally, \textit{DGCN} model leverages graph convolutional networks for capturing collaborative effects in the user-item bipartite graph ensuring diverse recommendations through rebalanced neighbor discovery~\cite{zheng2021dgcn}.
    
    \item \textbf{Disentangling mechanisms~\footnote{Disentangling mechanisms aim to separate and capture distinct factors or patterns within graph data, ensuring more interpretable and robust recommendations by reducing the entanglement of various latent factors~\cite{ma2019disentangled}.}:} \textit{DGCF} framework diversifies recommendations by disentangling user intents in collaborative filtering using intent-aware graphs and a graph disentangling layer~\cite{wang2020disentangled}.

    \item \textbf{Dynamic graph construction~\footnote{Dynamic graph construction involves continuously updating and evolving the graph structure to incorporate new interactions and/or entities~\cite{skarding2021foundations}.}:} \textit{DDGraph} approach involves dynamically constructing a user-item graph to capture both user-item interactions and non-interactions, and then applying a novel candidate item selection operator to choose items from different sub-regions based on distance metrics~\cite{ye2021dynamic}.
    
    \item \textbf{Adversarial learning~\footnote{Adversarial examples in recommender systems, as a form of data augmentation, bolster data diversity for improved generalization, counteract inherent biases, and ensure fair node representation in GNNs for fairer recommendations~\cite{deldjoo2021survey}.}:} To improve the accuracy-diversity trade-off in tag-aware systems, the \textit{DTGCF} model utilizes personalized category-boosted negative sampling, adversarial learning for category-free embeddings, and specialized regularization techniques~\cite{zuo2023dtgcf}. Furthermore, the above-mentioned \textit{DGCN} model also employs adversarial learning to make item representations more category-independent.
    
    \item \textbf{Contrastive learning~\footnote{Contrastive learning pushes similar item or user embeddings closer and dissimilar ones apart to enhance recommendation quality~\cite{liu2021contrastive}.}:} The Contrastive Co-training (\textit{CCT}) method by~\cite{ma2022contrastive} employs an iterative pipeline that augments recommendation and contrastive graph views with pseudo edges, leveraging diversified contrastive learning to address popularity and category biases in recommendations.
    

    \item \textbf{Heterogeneous Graph Neural Networks~\footnote{Heterogeneous graph neural networks process diverse types of nodes and edges, capturing complex relationships using a heterogeneous graph as input~\cite{wu2022graph}.}:} The \textit{GraphDR} approach by \cite{xie2021improving} utilizes a heterogeneous graph neural network, capturing diverse interactions and prioritizing diversity in the matching module.
\end{itemize}

Each of these methods offers a unique approach to the accuracy-diversity challenge. While all aim to improve the trade-off, their strategies vary, highlighting the multifaceted nature of the challenge at hand.

\section{Serendipity in GNN-based Recommender Systems}

\subsection{Definition and importance of serendipity and novelty}

Serendipity and closely related novelty are crucial in recommender systems, both aiming to boost user discovery. Serendipity refers to surprising yet relevant recommendations, promoting exploration and curiosity. Novelty suggests new or unfamiliar items, expanding user exposure. Both prevent over-specialization and encourage user curiosity~\cite{kaminskas2016diversity}.


\subsection{Review of recent developments in promoting serendipity and novelty}
Recent advancements in GNN-based recommender systems have shown promising results in promoting serendipity and novelty, although notably fewer efforts have been directed towards balancing the accuracy-serendipity and accuracy-novelty trade-offs in comparison to the accuracy-diversity trade-off. In our exploration, we identified several studies addressing these efforts and have categorized them based on the primary theme of their contribution:

\begin{itemize}
    \item \textbf{Neighbor-based mechanisms:} Approach proposed by~\cite{boo2023serendipity} enhances session-based recommendations by incorporating serendipitous session embeddings, leveraging session data and user preferences to amplify global embedding effects enabling users to control explore-exploit tradeoffs.

    \item \textbf{Long-tail recommendations~\footnote{Long-tail recommendations focus on suggesting less popular or niche items~\cite{kowald2020unfairness}.}:} The \textit{TailNet} architecture is designed to enhance long-tail recommendation performance. It classifies items into short-head and long-tail based on click frequency and integrates a unique preference mechanism to balance between recommending niche items for serendipity and maintaining overall accuracy~\cite{liu2020long}.

    \item \textbf{Normalization techniques~\footnote{Normalization techniques in GNN-based recommender systems stabilize and scale node features or edge weights, ensuring consistent and improved model convergence and recommendation quality~\cite{gupta2019niser}.}:}~\cite{zhao2022investigating} proposed \textit{r-AdjNorm}, a simple and effective GNN improvement that can improve the accuracy-novelty trade-off by controlling the normalization strength in the neighborhood aggregation process.
    
    \item \textbf{General GNN architecture enhancements~\footnote{We refer to general GNN architecture enhancements in recommender systems as the advancements in architectures, aggregators, or training procedures that better capture graph structures for improved recommendation accuracy.}:} Similarly to the popular \textit{LightGCN} approach by~\cite{he2020lightgcn}, the \textit{ImprovedGCN} model by~\cite{dhawan2022improvedgcn} adapts and simplifies the graph convolution process in GCNs for item recommendation, inadvertently boosting serendipity. On the other hand, the \textit{BGCF} framework by~\cite{sun2020framework}, designed for diverse and accurate recommendations, also boosts serendipity and novelty through its joint training approach. These GNN-based models, while focusing on accuracy, inadvertently elevate recommendation serendipity and/or novelty. 
\end{itemize}

These studies collectively demonstrate the potential of GNNs in enhancing the serendipity and novelty of recommender systems, while also highlighting the need for further research to address existing challenges.

\section{Fairness in GNN-based Recommender Systems}
\subsection{Definition and importance of fairness}

Fairness in recommender systems ensures no bias towards certain users or items. It can be divided into user fairness, which avoids algorithmic bias among users or demographics, and item fairness, which ensures equal exposure for items, countering popularity bias~\cite{leonhardt2018user,kowald2020unfairness,abdollahpouri2021user,kowald2022drives,ecir_bias_2023,lex2020modeling}. Fairness helps to mitigate bias, supports diversity, and boosts user satisfaction. In GNN-based systems, which can amplify bias, fairness is crucial for balanced recommendations and optimal performance~\cite{ekstrand2018exploring,chizari2022comparative,chen2023graph,gao2023survey}.


\subsection{Review of recent developments in promoting fairness}


In the evolving landscape of GNN-based recommender systems, the pursuit of user and item fairness has become a prominent topic. Recent advancements can be broadly categorized based on the thematic emphasis of their contributions:

\begin{itemize}
    \item \textbf{Neighbor-based mechanisms:} The \textit{Navip} method debiases the neighbor aggregation process in GNNs using "neighbor aggregation via inverse propensity", focusing on user fairness~\cite{kim2022debiasing}. Additionally, the \textit{UGRec} framework by \cite{liu2022mitigating} employs an information aggregation component and a multihop mechanism to aggregate information from users' higher-order neighbors, ensuring user fairness by considering male and female discrimination. The \textit{SKIPHOP} approach focuses on user fairness by introducing an approach that captures both direct user-item interactions and latent knowledge graph interests, capturing both first-order and second-order proximity. Using fairness for regularization, it ensures balanced recommendations for users with similar profiles~\cite{wu2022equipping}.

    \item \textbf{Multimodal feature learning~\footnote{Multimodal feature learning integrates diverse data sources, like text, images, and graphs, into unified embeddings to enrich recommendation context and accuracy~\cite{zhou2023comprehensive}.}:} The method proposed by~\cite{li2019long} fuses hashtag embeddings with multi-modal features, considering interactions among users, micro-videos, and hashtags.

    \item \textbf{Adversarial learning:} The \textit{UGRec} model additionally incorporates adversarial learning to eliminate gender-specific features while preserving common features.

    \item \textbf{Contrastive learning:} 
    The \textit{DCRec} model by \cite{yang2023debiased} leverages debiased contrastive learning to counteract popularity bias and addressing the challenge of disentangling user conformity from genuine interest, focusing on user fairness. The \textit{TAGCL} framework also capitalizes on the contrastive learning paradigm, ensuring item fairness by reducing biases in social tagging systems~\cite{xu2023fairness}.

    \item \textbf{Long-tail recommendations:} The \textit{NISER} method by \cite{gupta2019niser} addresses the long-tail issue by focusing on popularity bias in session-based recommendation systems. It aims to ensure item fairness by normalizing item and session representations, thereby improving recommendations, especially for less popular items. Additionally, the above-mentioned approach by~\cite{li2019long} also focuses on long-tail recommendations.

    \item \textbf{Self-training mechanisms~\footnote{Self-training mechanisms leverage unlabeled data by iteratively predicting and refining labels, enhancing the model's performance with augmented training data.~\cite{yu2023self}}:} The \textit{Self-Fair} approach by \cite{liu2022self} employs a self-training mechanism using unlabeled data with the goal of improving user fairness in recommendations for users of different genders. By iteratively refining predictions as pseudo-labels and incorporating fairness constraints, the model balances accuracy and fairness without relying heavily on labeled data.
\end{itemize}

In the broader context of graph neural networks, researchers have also tackled fairness in non-recommender systems tasks, such as classification~\cite{dai2021say,ma2021subgroup,dong2022edits,zhang2022trustworthy}. Their insights provide valuable lessons for future development of fair recommender systems.

\section{Discussion and Future Directions}

In this paper, we have conducted a comprehensive review of the literature on diversity, serendipity, and fairness in GNN-based recommender systems, with a focus on optimizing beyond-accuracy metrics. Throughout our analysis, we have explored various aspects of model development and discussed recent advancements in addressing these dimensions.  

To further advance the field and guide future research, we have formulated three key questions:

\emph{Q1: What are the practical challenges in optimizing GNN-based recommender systems for beyond-accuracy metrics?}


GNNs are able to capture complex relationships within graph structures. However, this sophistication can lead to overfitting, especially when prioritizing accuracy~\cite{fu2023deep}. Data sparsity and the need for auxiliary data, such as demographic information, challenge the optimization of high-quality node representations, introducing biases~\cite{dhawan2022improvedgcn}. An overemphasis on past preferences can limit novel discoveries~\cite{dhawan2022improvedgcn}, and while addressing popularity bias is essential, it might inadvertently inject noise, reducing accuracy~\cite{liu2020long}. Balancing diverse objectives, like fairness, accuracy, and diversity, is nuanced, especially when optimizing one can compromise another~\cite{liu2022mitigating}. These challenges emphasize the need for focused research on effective modeling of GNN-based recommender systems focused on beyond-accuracy optimization.

\emph{Q2: Which model development stages of GNN-based recommender systems have seen the most innovation for tackling beyond-accuracy optimization, and which stages have been underutilized?}

By conducting a thorough analysis of the reviewed papers (see Table~\ref{table:summary}), we have observed that the graph construction, propagation layer, and training methodologies have seen significant innovation in GNN-based recommender systems. This includes advanced graph construction methods, innovative graph convolution operations, and unique training methodologies. However, stages like embedding initialization, embedding fusion, and score computation are relatively underutilized. These stages could offer potential avenues for future research and could provide novel ways to balance accuracy, fairness, diversity, novelty, and serendipity in recommendations.

\emph{Q3: What are potentially unexplored areas of beyond-accuracy optimization in GNN-based recommender systems?}

A less explored aspect in GNN-based recommender systems is personalized diversity, which modifies the diversity in recommendations to match individual user preferences. Users favoring more diversity get more diverse recommendations, whereas those liking less diversity get less diverse ones~\cite{eskandanian2017clustering}. This concept of personalized diversity, currently under-researched in GNN-based systems, hints at an intriguing future research direction. It can also relate to personalized serendipity or novelty, tailoring unexpected or novel recommendations to user preferences. Thus, incorporating personalized diversity, serendipity, and novelty in GNN-based systems could enrich beyond-accuracy optimization.

Overall, this review aims to help researchers and practitioners gain a deeper understanding of the multifaceted issues and potential avenues for future research in optimizing GNN-based recommender systems beyond traditional accuracy-centric approaches. By addressing the practical challenges, identifying underutilized model development stages, and highlighting unexplored areas of optimization, we hope to contribute to the development of more robust, diverse, serendipitous, and fair recommender systems that cater to the evolving needs and expectations of users.


\section*{Conflict of Interest Statement}

The authors declare that the research was conducted in the absence of any commercial or financial relationships that could be construed as a potential conflict of interest.

\section*{Author Contributions}
TD: literature analysis, conceptualization, writing. ELA: conceptualization, writing. ELE and DK: conceptualization, writing, supervision. All authors contributed to the article and approved the submitted version.

\section*{Acknowledgments}

This work is supported by the ``DDIA'' COMET Module within the COMET – Competence Centers for Excellent Technologies Programme, funded by the Austrian Federal Ministry for Transport, Innovation and Technology (bmvit), the Austrian Federal Ministry for Digital and Economic Affairs (bmdw), FFG, SFG, and partners from industry and academia. The COMET Programme is managed by FFG. This research received support by the TU Graz Open Access Publishing Fund. Additional credit is given to OpenAI for the generative AI models, GPT-4 and ChatGPT, used in this work for text summarization, and sentence rephrasing. Verification of accuracy and originality was performed for all content generated by these tools.

\bibliographystyle{Frontiers-Vancouver} 
\bibliography{bibliography}


\end{document}